\documentclass[a4paper,preprint, prl,showpacs]{revtex4}
\usepackage{amsmath,amssymb}
\usepackage{makeidx}
\usepackage{amsfonts}
\usepackage[ansinew]{inputenc}
\usepackage[usenames,dvipsnames]{pstricks}
\usepackage{epsfig,graphicx}

\newcommand{\rme}{\mathrm{e}}

\newcommand{\rmd}{\mathrm{d}}

\makeindex

\begin{document}

\title{Projected generalized free energies for non-equilibrium states}

\author{Alexander Mozeika}

\affiliation{Department of Information and Computer Science,Aalto University, FI-00076, Aalto, Finland.}

\date{\today}

\begin{abstract}
We develop a systematic procedure to approximate the generalized free energy in out of equilibrium stochastic systems. The procedure only requires knowledge of the averages of macroscopic observables and uses a quasi-equilibrium distribution to this task. As an application we consider model systems in the regime of diverging relaxation times. We find that the geometry of the approximate generalized free energy  changes at the onset of this phenomenon.
\end{abstract}
\maketitle
Equilibrium is one of the fundamental assumptions in Statistical Physics (SP): a system without any external disturbances tends to a state where macroscopic quantities such as pressure in gas,  magnetization in magnetic alloys, etc., fluctuate, but on the average remain unchanged~\cite{LandauLifshitz1969}. A system in equilibrium is governed by the Gibbs-Boltzmann distribution $p_{eq}(s)=\rme^{\frac{1}{k_BT}(F[p_{eq}] - E(s))}$, where $E(s)$ is the energy of a \emph{microstate} $s$, $T$ is the temperature and  $k_B$ the Boltzmann constant (we set $k_B=1$ for convenience). The averages of functions of microstates (\emph{observables}) in such a system are related via various thermodynamic functions~\cite{LandauLifshitz1969}. One of these functions is the free energy $F[p_{eq}]=E[p_{eq}]-TS[p_{eq}]$,  where $E[p_{eq}]=\sum_s p_{eq}(s)E(s)$ is the average energy and $S[p_{eq}]=-\sum_s p_{eq}(s)\log p_{eq}(s)$ is the Gibbs-Shanon entropy. Computation of   $F[p_{eq}]$ for many models of realistic interacting systems is very difficult. To circumvent this problem, one usually replaces  $p_{eq}(s)$ with a simpler distribution, requiring this distribution to be, in some sense, as close as possible to $p_{eq}(s)$. This procedure can be seen as a projection of the potentially infinite-dimensional~\cite{Sollich} free energy into some lower-dimensional space.  

This situation in equilibrium SP is repeating itself in a much more complex non-equilibrium scenario. There the system is usually described by the probability distribution $p(s)$ which, in contrast to $p_{eq}(s)$, also has an additional dimension, time. To follow the evolution of $p(s)$ directly for any interacting system of reasonable size is usually not an option.  In the past this problem was approached by choosing a finite set of macroscopic observables $\Omega(s)=(\Omega_1(s),\ldots,\Omega_L(s))$,  then trying to derive a closed system of dynamical equations for their averages $\Omega=(\Omega_1,\ldots,\Omega_L)$, i.e. $\Omega_{\mu}=\sum_s p(s)\Omega_{\mu}(s)$~\cite{Zubarev,Zwanzig}. The method of non-equilibrium statistical operator~\cite{Zubarev} uses a so-called quasi-equilibrium distribution $\hat p(s)$, which can be seen as a generalization of $p_{eq}(s)$, for this task. Then the requirement that $\hat p(s)$ has to be consistent with $\Omega$ gives rise to their conjugates  $\hat\Omega$ (akin to $E_{eq}$ and $T$). This framework allows to define thermodynamic functions but the meaning of these functions in a more general non-equilibrium setting is not clear. 
 
Recent analytic approaches to this problem~\cite{DRT} which are reminiscent of~\cite{Zubarev} but are specifically tailored to deal with the systems with quenched disorder~\cite{SGbook}, have enjoyed a degree of success in the non-equilibrium SP of disordered systems ~\cite{DRTforSK,  DRTforFC, Mozeika}.  The main motivation for this direction of research  was the difficulty of implementing  exact analytical methods~\cite{Dominicis}, which are very successful for some important models ~\cite{Kurchan}, but in general are very difficult to implement~\cite{Hatchett, SemerjianCugliandoloMontanari} and  also unable to describe one-time quantities~\cite{Caltagirone}. The thermodynamic functions were usually not discussed in these works.  

Formally the definition $F[p_{eq}]=E[p_{eq}]-TS[p_{eq}]$ can be generalized by replacing $p_{eq}(s)$ with $p(s)$, giving rise to the generalized free energy $F[p]=E[p]-TS[p]$. In discrete state systems, which tend to equilibrium,  $F[p]$  is monotonically decreasing with time and is bounded below by $F[p_{eq}]$~\cite{vanKampen}  (to show this, one usually studies how the relative entropy $D(p\vert\vert p_{eq})=\sum_s p(s)\log\frac{p(s)}{p_{eq}(s)}=(F[p] -F[p_{eq}])/T\geq0$ is evolving in time). Thus $F[p]$ is a Lyapunov function for these systems. In fact this is true for any system (also with continuous  $s$) tending to $p_{eq}(s)$ where $p(s)$ is governed by a linear operator~\cite{Voigt}. Remarkably, for non-equilibrium systems which are close to equilibrium, $F[p]$ is related to a measurable thermodynamic work~\cite{Crooks, Sasa}.

In this Letter, we develop a method which allows a systematic  way to approximate $F[p]$. This procedure can be seen as a projection of $F[p]$ onto some finite set of average macroscopic quantities $\Omega$, giving rise to $F[\hat p]$. The method is very general and allows to study the evolution of $F[\hat p]$ with time  in systems which tend to thermal equilibrium.  We apply this method to a model system with slow dynamics. In a simple variant of this model we rigorously show that $F[\hat p]$ is a Lyapunov function for the dynamics of $\Omega$. For a more complicated version of the same model we are only able to check this numerically. In both cases change in the shape of the surface of $F[\hat p]$ signals onset of the slow dynamics.

We consider a system  of $N$ Ising spins $s_i\in\{-1,1\}$ interacting on a graph (generalization to other discrete-state  systems governed by master equations is straight-forward). The evolution of the microscopic state  $s\in\{-1,1\}^N$ is stochastic and is governed by the master equation  
\begin{eqnarray}
  \frac{\rmd}{\rmd t} p(s) & = & \sum_{i=1}^N [p(F_i s)w_i(F_i s)-p(s)w_i(s)],\label{eq:master}%\sum_{\tilde s}\mathrm{W}[\;s\;\vert\;\tilde s\;]\; p_t(\tilde s)
\end{eqnarray}
where $F_i$ is a flip operator, i.e. $f(F_is)=f(s_1,\ldots,-s_i,\ldots,s_N)$, and   $w_i(s)=\frac{1}{2}[1-s_i \tanh[\beta h_i(s)]]$ is a Glauber transition rate. The choice for $h_i(s)=-\frac{1}{2}\sum_{s_i}E(s)s_i$ (field) ensures that process (\ref{eq:master}) evolves towards the equilibrium distribution $p_{eq}(s)$ with $T=1/\beta$~\cite{longpaper}. The properties of observables $\Omega (s)$ are fully described by the  probability distribution $P(\Omega)=\sum_{s}p(s)\prod_\mu\delta[\Omega_\mu -\Omega_\mu(s)]$. For $N\rightarrow\infty$ the distribution  $P(\Omega)$ has a deterministic evolution~\cite{DRT}  
\begin{eqnarray}
  \frac{\rmd}{\rmd t}\Omega =\sum_{s}P(s\vert \Omega)  \sum_{i=1}^N w_i(s)\left[\Omega(F_i s)-\Omega(s)\right],\label{eq:observ}
\end{eqnarray}
where
\begin{eqnarray}
P(s\vert \Omega)  & = &
\frac{p(s)\prod_\mu\delta \left[\Omega_\mu - \Omega_\mu (s)\right]}{\sum_{\tilde s}p(\tilde s)\prod_\mu\delta \left[\Omega_\mu - \Omega_\mu (\tilde s)\right]}\label{def:OmegaAver}.
\end{eqnarray}
Equation (\ref{eq:observ}) is exact but not closed due to the presence of $p(s)$. In order to close this equation one usually assumes \emph{equi-partitioning}: $p(s)$ depends on $s$ only via $\Omega(s)$~\cite{DRT}. This gives rise to the micro-canonical distribution
\begin{eqnarray}
p_{\Omega}(s) & = &
\frac{\prod_\mu\delta \left[\Omega_\mu - \Omega_\mu (s)\right]}{\sum_{\tilde s}\prod_\mu\delta \left[\Omega_\mu - \Omega_\mu (\tilde s)\right]}\label{def:micro}
\end{eqnarray}
which replaces $P(s\vert \Omega)$ in equation (\ref{eq:observ}) and thereby closes this system of equations. This new system of equations is no longer exact but constitutes an \emph{approximation} to the true dynamics of $\Omega$. The quality of this approximation depends crucially on the choice of $\Omega (s)$. For systems tending to  $p_{eq}(s)$, the set $\Omega (s)$ should at least contain the energy $E(s)$  and some function $g(s)$ that specifies initial conditions, i.e. $p(s)\equiv p(g(s))$ at $t=0$~\cite{DRT}. Furthermore, there is a systematic way of improving this approximation~\cite{ SemerjianWeigt, Mozeika}.

Assuming equivalence of micro-canonical and canonical ensembles, which is expected for $N\rightarrow\infty$, leads us to the canonical distribution   
\begin{eqnarray}
\hat{ p}(s) & = & \exp\bigg[\sum_\mu\hat\Omega_\mu\,\Omega_\mu (s) +\Phi\bigg]\label{eq:canon}
\end{eqnarray}
where $\Phi=-\log\sum_{\tilde s}\exp[  \sum_\mu        \hat\Omega_\mu\,\Omega_\mu (\tilde s)]$. Using this distribution  is more convenient than  (\ref{def:micro}),  and  it establishes a connection with~\cite{Zubarev}. In this framework,  given $\Omega$, at any time $t$ the variables $\hat\Omega$ are obtained by minimizing the potential 
\begin{eqnarray}
G & = &\Phi +  \sum_\mu\hat\Omega_\mu\,\Omega_\mu.\label{eq:gibbs}
\end{eqnarray}
The relations
\begin{eqnarray}
\Omega_\mu = -\frac{\partial }{\partial \hat\Omega_\mu}\Phi,\,\,\hat\Omega_\mu =  \frac{\partial }{\partial\Omega_\mu}G \label{eq:thermo}
\end{eqnarray}
identify $\Omega$ and $\hat\Omega$ as conjugate variables~\cite{Zubarev}.

Furthermore, the (projected) non-equilibrium free energy associated with the whole scheme is given by 
\begin{eqnarray}
F[\hat{ p}] = E[\hat{ p}] -T S[\hat{ p}] \label{eq:F}
\end{eqnarray}
where $S[\hat{ p}]=- G$ is a Gibbs-Shanon entropy and $E[\hat{ p}]= \sum_s\hat{ p}(s) E(s)$. From this, it follows that $F[\hat{ p}]$ evolves in time according to the equation  
\begin{eqnarray}
\frac{\rmd}{\rmd t}F[\hat{ p}] =\frac{\rmd}{\rmd t} E[\hat{ p}] +     T \sum_\mu\hat\Omega_\mu\,\frac{\rmd \Omega_\mu}{\rmd t}.    \label{eq:dFdt}
\end{eqnarray}
The relation between $F[\hat{ p}]$ and $F[ p]$ of the \emph{true} dynamics is simple when $\Omega_\mu= \sum_s\hat{ p}(s)\Omega_\mu (s)=\sum_s p(s)\Omega _\mu(s)$. Then  $D(p\vert\vert \hat p)= S[\hat{ p}]-  S[p]\geq0$ and we have the inequality  
\begin{eqnarray}
F[p]\geq   F[\hat{ p}]\geq  F[p_{eq}]\label{eq:ineq1}
\end{eqnarray}
where we  have used  $D(\hat p\vert\vert p_{eq}) \geq0$ to obtain the second inequality. The relation is more complicated when we have deviations from the true averages  $\sum_s p(s)\Omega _\mu(s)$. Suppose that this deviation is given by the difference $\Delta\Omega_\mu=\sum_s[\hat{ p}(s)-p(s)]\Omega _\mu(s)$. Then again using $D(p\vert\vert \hat p)\geq0$ we obtain the following inequality  
\begin{eqnarray}
F[p]-   F[\hat{ p}]\geq  E[p]-E[\hat{ p}]  -T \sum_\mu\hat\Omega_\mu\,\Delta\Omega_\mu. \label{eq:ineq2}
\end{eqnarray}

Let us now consider systems with the energy function 
\begin{eqnarray}
E(s)=-J\sum_{<i_1,\ldots,i_p>}s_{i_1}s_{i_2}\ldots s_{i_p},\label{def:p-spin-E}
\end{eqnarray}
where the sum is over all edges of some (hyper) graph. The choice of (\ref{def:p-spin-E}) is motivated by the presence of metastable states~\cite{Binder} in this model for $p\geq 3$, which drastically affect the dynamics~\cite{MeltingDyn, MeltingDyn2}.  Firstly, we consider a fully-connected variant of (\ref{def:p-spin-E}) where $E(s)=-Nm^p(s) $ with $m(s)=\frac{1}{N}\sum_{i=1}^N s_i$ (magnetization). It was argued in~\cite{MeltingDyn} that this simple model  has features which are usually found in super-cooled liquids.

\begin{figure}[t]
\vspace*{0mm} \hspace*{-0mm} \setlength{\unitlength}{0.27mm}
\begin{picture}(350,105)
%\put(157,110){\includegraphics[height=100\unitlength,width=160\unitlength]{fig1b.eps}}
%\put(0,110){\includegraphics[height=100\unitlength,width=160\unitlength]{fig1a.eps}}
\put(159,0){\includegraphics[height=100\unitlength,width=160\unitlength]{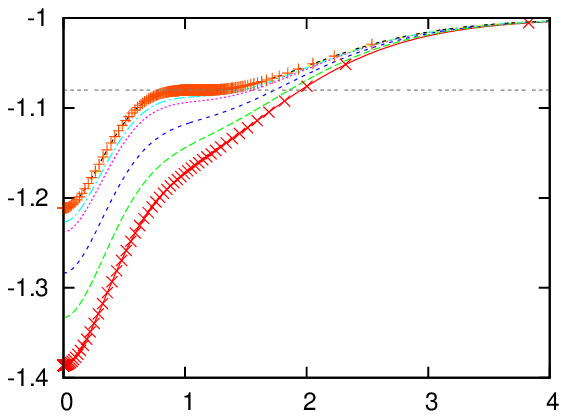}}
\put(245,-7){$\hat{m}$} %\put(245,20){\scriptsize{$h\!=\!0,k\!=\!2$}}
\put(0,0){\includegraphics[height=100\unitlength,width=160\unitlength]{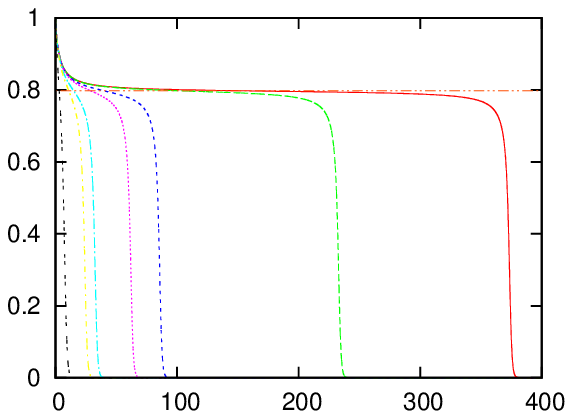}}
\put(-3,65){$m$}
\put(85,-7){$t$}  \put(160,65){$f$}
\put(130,80){$(a)$}\put(290,80){$(b)$}
\end{picture}
 \vspace*{0mm}
\caption{(Color online). Evolution of the magnetization $m$ and the generalized free energy  $f$ with time $t$ from the  fully magnetized state $m=1$ in a fully-connected $3$-spin Ising model:  (a) Evolution of $m$ for $\beta \in [0.5,0.5721]$  (from left to right). The horizontal line corresponds to the value of $m=0.7979$ at $\beta_s=0.5722$.  (b) Evolution of $f$ (symbols) on its surface (lines) as a function of $\hat{m}$ for $\beta \in [0.5,0.5721]$  (from bottom to top). The direction of time is from right to left. The horizontal line corresponds to the value of $f_{eq}= -1.0802$ at $\beta_s=0.5722$.\label{fig:1} \vspace*{-0.0cm}
}
\end{figure}

In equilibrium ($p=3$) and for high temperature the free energy density $f_{eq}=\lim_{N\rightarrow\infty}F[p_{eq}]/N$ has only one global minimum  corresponding to the \emph{paramagnetic} $m=0$ state. As we lower the temperature $f_{eq}$ develops a second (local) minimum at $\beta_s=0.5722$ which corresponds to the  \emph{ferromagnetic} $m\neq0$ state and becomes only globally stable at $\beta_c=0.6721$ inducing a first order transition. As we approach $\beta_s$  from above, the dynamics of $m$ (from the fully ordered $m=1$ state) exhibits a considerable slowing down (see Figure \ref{fig:1} (a)). 

Let us now consider the projected dynamics of this model. We choose  $\Omega(s)=(Nm(s))$, then $\hat{ p}(s)=\exp[N\hat{m}m(s) +\Phi] $, where  $\Phi=-N\log 2\cosh(\hat{m})$. Then from relation (\ref{eq:thermo}), it follows that $m=\tanh(\hat{m}) $. Using this result we obtain $S[\hat{ p}]=N[\log 2\cosh(  \tanh^{-1}(m)  )    - \tanh^{-1}(m) m]$. The average magnetization $m$ is governed by the equation $\rmd m/\rmd t = -m + \sum_s \hat{ p}(s)\tanh(\beta p m(s)^{p-1})$. For $N\rightarrow\infty$ this equations evaluates to $\rmd m/\rmd t = -m + \tanh(\beta p \,m^{p-1})$ which recovers the exact result for this model~\cite{MeltingDyn}.In this limit, $E=-Nm^p$ and hence we 
obtain $f(m)=\lim_{N\rightarrow\infty}F[\hat{ p}]/N = -m^p -T(\log 2\cosh(  \tanh^{-1}(m)  )    - \tanh^{-1}(m) m)$. The shape of this function is directly related to the slowing down in the dynamics of $m$ (see Figure \ref{fig:1} (b)).

Furthermore,  $f(m)$ is a Lyapunov function for the evolution of $m$. In order to show this we consider $\frac{\rmd f(m)}{\rmd t}=\frac{\partial f(m)}{\partial m}\frac{\rmd m}{\rmd t}  = -T[- \tanh^{-1}(m) + \beta pm^{p-1}][-m +         \tanh(\beta p \,m^{p-1}) ]$. Now either $m \geq \tanh(\beta p \,m^{p-1})$ or  $m < \tanh(\beta p \,m^{p-1})$ then, by using monotonicity of $\tanh^{-1}$, both terms in square brackets are either (simultaneously) positive or negative and hence $\frac{\rmd f(m)}{\rmd t}\leq 0$. Furthermore,  $f(m)$ is bounded from below by $f_{eq}=f(m_{eq})$, where $m_{eq}$ is the solution of $m = \tanh(\beta p \,m^{p-1})$ with the lowest $f(m)$~\cite{MeltingDyn}. 

Now we turn our attention to the more complicated 3-spin  variant of (\ref{def:p-spin-E}) defined on a random $k+1$-regular graph, i.e. the graph is finitely-connected.  This model was studied in the past, because some of its properties are reminiscent of those  found in structural glasses~\cite{3SpinFerro}. The thermodynamic behavior of this model  is similar in some aspects to the fully-connected model ($\beta_s=0.6135$ and $\beta_c=0.8264$ for this model) but for lower temperatures it is in a glass state~\cite{3SpinFerro}. 

\begin{figure}[t]
\vspace*{0mm} \hspace*{-0mm} \setlength{\unitlength}{0.27mm}
\begin{picture}(350,210)
\put(157,110){\includegraphics[height=100\unitlength,width=160\unitlength]{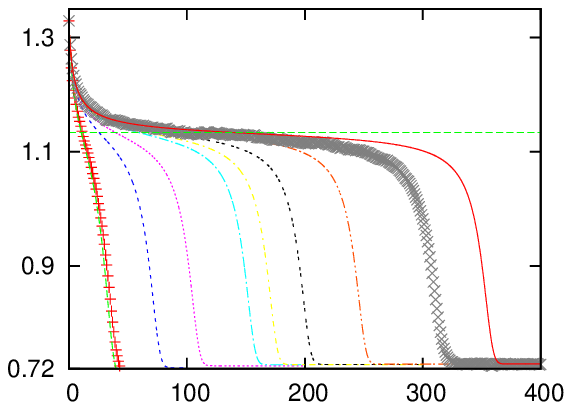}}
\put(0,110){\includegraphics[height=100\unitlength,width=160\unitlength]{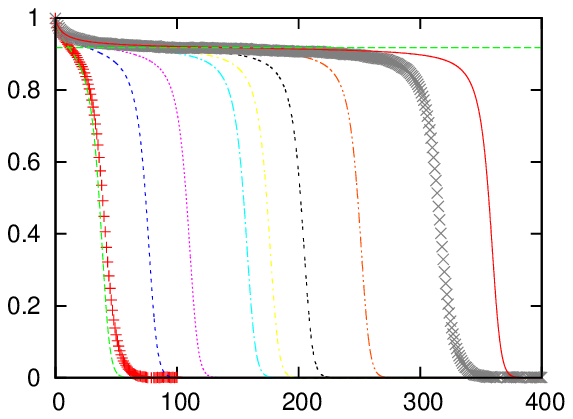}}
\put(145,-10){\includegraphics[height=130\unitlength,width=190\unitlength]{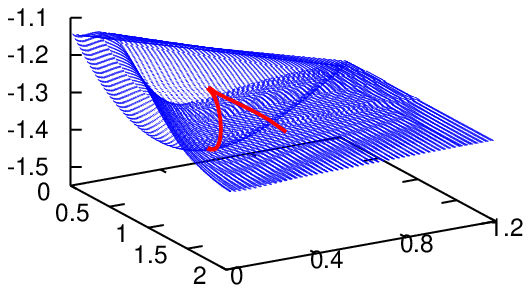}}
\put(180,13){$\hat{m}$} \put(265,3){$\hat{E}$}
\put(0,0){\includegraphics[height=100\unitlength,width=160\unitlength]{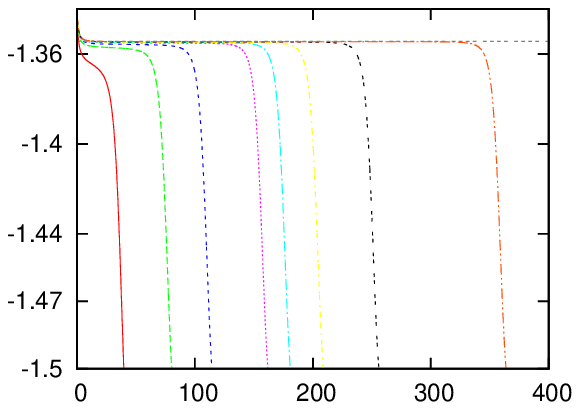}}
\put(-3,175){$m$} \put(-3,65){$f$}
\put(85,-7){$t$} \put(160,175){$-E$} %\put(160,65){$f$}
\put(130,190){$(a)$}\put(290,190){$(b)$}\put(130,80){$(c)$}\put(290,80){$(d)$}
\end{picture}
 \vspace*{0mm}
\caption{(Color online). Evolution of the magnetization $m$, energy $E$ and generalized free energy  $f$ with time $t$ from the  fully magnetized state $m=1$ in a finitely-connected $3$-spin Ising model with $k=3$. Theoretical results (lines) are plotted against the results of MC simulations (symbols) with $N=10^6$. Top: (a) Evolution of $m$ for $\beta \in [0.5882, 0.6131]$  (from left to right). The horizontal line corresponds to the value of $m=0.9179$ at $\beta_s=0.6135$; (b) Evolution of $E$ for the same $\beta$. The horizontal line corresponds to the value of $-E=1.1338$ at $\beta_s$. Bottom: (c) Evolution of $f$ for the same $\beta$. The horizontal line corresponds to the value of $f_{eq}= -1.3544$ at $\beta_s$; (d) Evolution of $f$ (solid line) on its surface as a function of $\hat{m}(t)$ and $\hat{E}(t)$ for $\beta =0.6131$. The direction of time is from right to left. The deepest point corresponds to $f_{eq}= -1.5160$. \label{fig:2} \vspace*{-0.0cm} }
\end{figure}

Let us start with the simplest approximation, which uses $\Omega(s)=(N m(s), E(s))$. We will call this the $(m,E)$-approximation. This approximation gives rise to  $\hat{ p}(s)=\exp[N\hat{m}m(s) -\hat{E}E(s)+\Phi]$. Solving the projected dynamics requires computation of $D(\sigma, h)=\lim_{N\rightarrow\infty}\sum_s\hat{ p}(s)D(\sigma, h; s)$, where $D(\sigma, h; s)=\frac{1}{N}\sum_{i=1}^N\delta_{\sigma,s_i}\delta\left(h-h_i(s)\right)$ (joint spin-field density),  with the cavity method~\cite{ DRTforFC, IPCbook}. This computation leads to the system of equations
\begin{eqnarray}
\frac{\rmd }{\rmd t}m &=& -m + \int\rmd h D(h)\tanh(\beta h) \label{eq:3-spin-observ}\\
\frac{\rmd}{\rmd t} E & =& -3E - \int \rmd h D(h)\tanh(\beta h) h  \nonumber\\
D(\sigma, h)   &= &\rme^{\sigma(\hat{E} h + \hat{m}                   )}\left\langle\delta(h - \sum_{a=1}^{k+1}s_1^a s_2^a )\right\rangle_{h_c},\nonumber
\end{eqnarray}
where $\left\langle(\ldots)\right\rangle_{h_c}=\sum_{\{s_j^a\}}\rme^{h_c\sum_{a=1}^{k+1}(s_1^a + s_2^a)+\mathcal{N}_D}(\ldots)$. The pair of conjugates $(\hat m,\hat E)$ and the cavity field $h_c$ are obtained from the equations
\begin{eqnarray}
m & = & \sum_\sigma D(\sigma)\sigma \label{eq:3-spin-inv}\\
E  &= &  - \frac{1}{3}\sum_\sigma \int\! \rmd h D(\sigma,h)\sigma h\nonumber\\ 
h_c&  =&  k\tanh^{-1}[\tanh(\hat{E})\tanh^2(h_c)] + \hat{m}.\nonumber
\end{eqnarray}
Close to $\beta_s=0.6135$ the $(m,E)$-approximation predicts a slowing down in the dynamics, as can be seen in Figures \ref{fig:2} (a) and (b). 
Furthermore, the cavity method~\cite{IPCbook} allows us to compute the entropy (density)   
\begin{eqnarray}
\frac{S[\hat{ p}]}{N}& = & -\frac{1}{3}(k\!+\!1)\!\!\!\sum_{s_0,\! s_1,\! s_2}\!\!\!P(s_0, s_1, s_2)\!\log\! P(s_0,\! s_1,\! s_2)\label{eq:3-spin-entropy}\\
&&+ k\sum_{s_0}P(s_0)\log P(s_0),\nonumber
\end{eqnarray}
where $P(s_0, s_1, s_2)=\rme^{\hat{E}s_0s_1s_2+h_c(s_0+ s_1+ s_2)+\mathcal{N}_P}$ and $P(s_0)$ is its marginal. The first contribution to this function is from the triangular plaquettes  and the second one from the single sites~\cite{IPCbook}. Using (\ref{eq:F}), result (\ref{eq:3-spin-entropy}) can be used to compute $f(\hat m, \hat E)=\lim_{N\rightarrow\infty}F[\hat{ p}]/N$. For the dynamics (\ref{eq:3-spin-observ}), $f(\hat m, \hat E)$  is monotonic decreasing with time for $\beta\leq\beta_s$, and develops a plateau  as we approach $\beta_s$ (see Figure \ref{fig:2} (c)). Furthermore, the surface of $f(\hat m, \hat E)$ becomes flat as we approach $\beta_s$ (see Figure \ref{fig:2} (d)).

In the $(m,E)$-approximation, equation (\ref{eq:dFdt}), when used for the density $f=\lim_{N\rightarrow\infty}F[\hat{ p}]/N$,  takes the simple form 
\begin{eqnarray}
T\frac{\rmd}{\rmd t}f =(\beta-\hat{E})\frac{\rmd}{\rmd t}E +\hat{m}\frac{\rmd}{\rmd t}m .\label{eq:dFdt2}
\end{eqnarray}
Although $f\geq f_{eq}$, for now we can verify that $\frac{\rmd}{\rmd t}f\leq0$ only numerically. 

\begin{figure}[t]
\vspace*{0mm} \hspace*{-0mm} \setlength{\unitlength}{0.27mm}
\begin{picture}(350,105)
%\put(157,110){\includegraphics[height=100\unitlength,width=160\unitlength]{fig1b.eps}}
%\put(0,110){\includegraphics[height=100\unitlength,width=160\unitlength]{fig1a.eps}}
\put(159,0){\includegraphics[height=100\unitlength,width=160\unitlength]{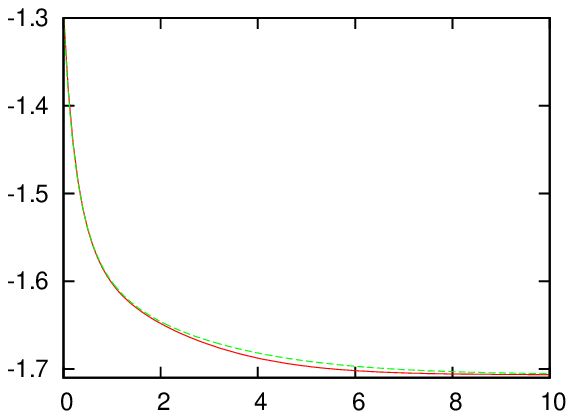}}
\put(245,-7){$t$} %\put(245,20){\scriptsize{$h\!=\!0,k\!=\!2$}}
\put(0,0){\includegraphics[height=100\unitlength,width=160\unitlength]{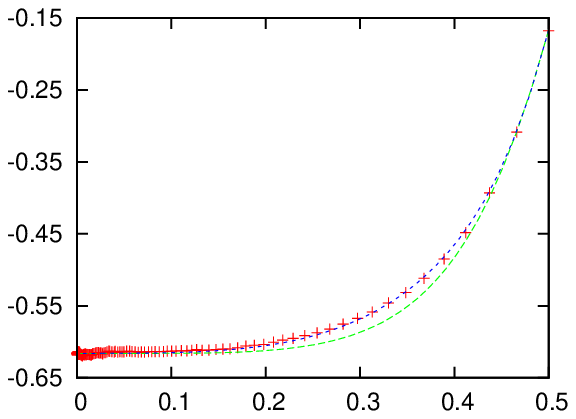}}
\put(-3,65){$E$}
\put(85,-7){$m$}  \put(160,65){$f$}
\put(130,80){$(a)$}\put(290,80){$(b)$}
\end{picture}
 \vspace*{0mm}
\caption{(Color online). Evolution of the magnetization $m$, energy $E$ and generalized free energy  $f$  with time $t$  in a finitely-connected $3$-spin Ising model for  $k=3$ and $\beta =0.5$. Theoretical results (lines) are plotted against the results of MC simulations (symbols) with $N=10^6$. (a) Evolution of $m$ and $E$ computed within  $(m, E)$-approximation (dashed line), $D(s,h)$-approximation (dotted line) and Monte Carlo simulation (symbols) of Glauber dynamics ($N=10^6$). The direction of time is from right to left. (b) Evolution of $f$ computed within the $(m, E)$-approximation (solid line) and the $D(s,h)$-approximation  (dashed line). \label{fig:3} \vspace*{-0.0cm}}
\end{figure}

Let us now consider the next level of approximation ($D(\sigma, h)$-approximation), in which we choose $\Omega(s)=N D(\sigma, h; s)$, leading us to the distribution $\hat{ p}(s)=\exp[  \sum_\sigma\int\rmd h \hat D(\sigma, h)D(\sigma, h; s) +\Phi]$.  The dynamic equation for $D(\sigma, h)$ requires the  computation of $A[\sigma,h;\tilde \sigma, \tilde h\vert F]=\lim_{N\rightarrow\infty}\sum_s\hat{ p}(s)\frac{1}{N}\sum_{<i,j>}\delta_{\sigma,s_j}\delta ( h -h_j
(F_is))\delta_{\tilde{\sigma},s_i}\delta(\tilde h -h_i (s))$ at every instance of time~\cite{longpaper}. In addition to this, we have to solve for $\hat D(\sigma,h )$ and the cavity function $Q(\sigma\vert\tilde\sigma)$. As can be seen in Figure \ref{fig:3}(a) the $D(\sigma, h)$-approximation can lead to significant improvements in the accuracy of predictions. It also  allows to study how $f\equiv f(\{\hat D(\sigma,h )\})$ evolves with time (see Figure \ref{fig:3}(b)) and its geometry. 

The generalized free energy is increasingly used in non-equilibrium statistical physics. The theoretical framework developed here provides a systematic way of approximating this high-dimensional object and its study offers a new geometrical  perspective on systems with slow dynamics. Furthermore, we envisage that it will offer us new insights into optimization problems which use thermal algorithms such as simulated annealing~\cite{Kirkpatrick}. 

\begin{acknowledgments}
 We thank E. Aurell, F. Ricci-Tersenghi, F. Krzakala,  R. Lemoy and C. Manzato for discussions and assistance. Support by the Academy of Finland (COIN, 251170) is acknowledged.
\end{acknowledgments}

%\bibliographystyle{apsrev}
%\bibliography{MaxEntr_refs}

\end{document}